\documentclass[12pt]{article}
\newlength{\medpaperheight}     
\newlength{\medpaperwidth}      
\newlength{\medtextheight}      
\newlength{\medtextwidth}       
\newlength{\medtopmargin}       
\newlength{\medoddsidemargin}   

\newcommand{\diag}{{\rm diag}}

\newcommand{\alf}{\alpha}

\newcommand{\del}{\delta}
\newcommand{\sig}{\sigma}

\newcommand{\Del}{\Delta}

\newcommand{\norm}[1]{ \left\| #1 \right\| }

\newcommand{\abs}[1]{\left|#1\right|}

\newcommand{\rhoh}{\hat{\rho}}

\newcommand{\Ccal}{{\mathcal C}}

\newcommand{\Ecal}{{\mathcal E}}

\newcommand{\eg}{\emph{e.g.}}
\newcommand{\ie}{\emph{i.e.}}

\newcommand{\bquem}{\begin{quote}\begin{em}}
\newcommand{\equem}{\end{em}\end{quote}}

\newcommand{\blist}{\begin{description}}
\newcommand{\elist}{\end{description}}

\newcommand{\bquote}{\begin{quote}}
\newcommand{\equote}{\end{quote}}

\newcommand{\ben}{\begin{enumerate}}
\newcommand{\een}{\end{enumerate}}

\newcommand{\bit}{\begin{itemize}}
\newcommand{\eit}{\end{itemize}}

\newcommand{\bea}{\begin{array}}
\newcommand{\eea}{\end{array}}

\newcommand{\bds}{\begin{displaystyle}}
\newcommand{\eds}{\end{displaystyle}}

\newcommand{\Rbf}{{\mathbf R}}
\newcommand{\Cbf}{{\mathbf C}}

\newcommand{\ds}{\displaystyle}

\newcommand{\mcal}[1]{ {\mathcal #1} }
\newcommand{\mbf}[1]{\mbox{\boldmath $#1$}}

\newcommand{\refeq}[1]{(\ref{eq:#1})}

\newcommand{\set}[2]{ \left\{ \,#1\, \left| \,#2\, \right.\right\} }

\newcommand{\seq}[1]{ \left\{ #1 \right\} }

\newcommand{\opt}{{\rm opt}}

\makeatletter
\def\beq{\@ifnextchar 
[{\@tempswatrue\@beq}{\@tempswafalse\@beq[]}}
\def\@beq[#1]{\begin{equation}\edef\@tmparg{#1}\ifx\@tmparg\@e
mpty \else
	\label{#1}\fi}
\newcommand{\eeq}{\end{equation}}
\makeatother 

\newcommand{\beqaa}{\begin{eqnarray*}}
\newcommand{\eeqaa}{\end{eqnarray*}}

\newcommand{\beqa}{\begin{eqnarray}}
\newcommand{\eeqa}{\end{eqnarray}}

\newcommand{\bc}{\begin{center}}
\newcommand{\ec}{\end{center}}

\usepackage{epsfig}
\usepackage{times}
\usepackage{amsmath,amssymb}

\usepackage{psfrag}

\newcommand{\udes}{L}

\newcommand{\rlxopt}{{\rm rlx\_opt}}

\newcommand{\Cbfsc}{\Cbf^{n_S \times n_C}}
\newcommand{\Cbfcs}{\Cbf^{n_C \times n_S}}
\newcommand{\Cbfscsc}{\Cbf^{n_S n_C \times n_S n_C}}
\newcommand{\Cbfcc}{\Cbf^{n_C \times n_C}}
\newcommand{\Cbfqq}{\Cbf^{n_S \times n_S}}

\newcommand{\fpur}{f_{\rm pure}}
\newcommand{\fmix}{f_{\rm mixed}}
\newcommand{\favg}{f_{\rm avg}}

\newcommand{\adj}{{\dagger}}

\newcommand{\btab}{\begin{tabular}}
\newcommand{\etab}{\end{tabular}}

\newcommand{\ket}[1]{\mbf{|}#1\mbf{\rangle}}
\newcommand{\bra}[1]{\mbf{\langle}#1\mbf{|}}

\newcommand{\Cbfmm}{\Cbf^{m\times m}}

\newcommand{\Rcal}{{\cal R}}
\newcommand{\Scal}{{\cal S}}

\newcommand{\Cbfnn}{{\bf C}^{n\times n}}

\newcommand{\trace}{{\bf Tr}}

\begin{document}

\title{
Quantum Error Correction
via Convex Optimization
}
\author{
Robert L. Kosut
\\
SC Solutions,
Sunnyvale, CA 94085
\\
kosut@scsolutions.com
\and
Daniel A. Lidar
\\
Depts. of Chemistry and Electrical Engineering-Systems
\\
University of Southern California,
Los Angeles, CA 90089
\\
lidar@usc.edu 
}

\date{}

\maketitle

\begin{abstract}

We show that the problem of designing a quantum information error
correcting procedure can be cast as a bi-convex optimization problem,
iterating between encoding and recovery, each being a semidefinite
program. For a given encoding operator the problem is convex in the
recovery operator. For a given method of recovery, the problem is
convex in the encoding scheme.  This allows us to derive new codes
that are locally optimal. We present examples of such codes that can
handle errors which are too strong for codes derived by analogy to
classical error correction techniques.

\end{abstract}



\section{Introduction}

Quantum error correction is essential for the scale-up of quantum
information devices. A theory of quantum error correcting codes has
been developed, in analogy to classical coding for noisy channels,
\eg, \cite{Shor:95,Steane:96,Gott:96,KnillL:97}.  Recently
\cite{ReimpellW:05} and \cite{YamamotoHT:05} did this by posing error
correction design as an optimization problem with the design variables
being the process matrices associated with the encoding and/or
recovery channels. Using fidelity measures leads naturally to a convex
optimization problem, specifically a semidefinite program (SDP)
\cite{BoydV:04}. The advantage of this approach is that noisy channels
which do not satisfy the standard assumptions for perfect correction
can be optimized for the best possible encoding and/or recovery.

In \cite{ReimpellW:05} the power-iteration method was used to find
optimal codes for various noisy channels, by alternately optimizing
the encoding and recovery channels. In contrast, here we apply convex
optimization via SDP, and similarly iterate between encoding and
recovery. For a given encoding operator the problem is convex in the
recovery. For a given method of recovery, the problem is convex in the
encoding. We further make use of Lagrange Duality to alleviate some of
the computational burden associated with solving the SDP for the
process matrices. The SDP formalism also allows for a robust design by
enumerating constraints associated with different error models.  We
illustrate the approach with an example where the error system does
not assume independent channels.

An intriguing prospect is to integrate the results found here within a
complete ``black-box'' error correction scheme, that takes quantum state
(or process) tomography as input and iterates until it finds an
optimal error correcting encoding and recovery.

\section{Quantum Error Correction}

\subsection{Standard model}

\begin{figure}[h]
\centering
\epsfig{file=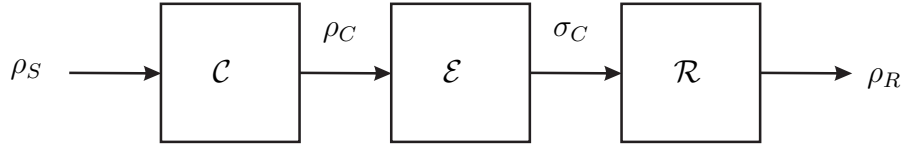,height=0.75in}
\caption{Standard encoding-error-recovery model of an error
correction system.}
\label{fig:rec}
\end{figure}

A {\em standard model} \cite[\S 10.3]{NielsenC:00} of an error
correction system as shown in the block diagram of Figure
\ref{fig:rec} is composed of three {\em quantum operations}: encoding
$\Ccal$, error $\Ecal$, and recovery $\Rcal$.
%

\psfrag{ccal}{$\Ccal$}
\psfrag{ecal}{$\Ecal$}
\psfrag{rcal}{$\Rcal$}
\psfrag{rhoc}{$\;\;\rho_{C}$}
\psfrag{sigc}{$\;\sig_{C}$}
\psfrag{psira}{$\ket{0_{RA}}$}
\psfrag{rhoq}{$\rho_S$}		
\psfrag{rhohq}{$\rho_R$}
\psfrag{ox}{}	
\psfrag{rhosab}{$\rho_{SAB}$}

The input, $\rho_S$, is the $n_S\times n_S$ dimensional density matrix
which contains the quantum information of interest and which is to be
processed. We will refer to $\rho_S$ as the {\em system state} or the
{\em unencoded state}. The output of the encoding operation is
$\rho_C$, the $n_C\times n_C$ dimensional {\em encoded state}. The
error operator, which is also the source of decoherence, corrupts the
encoded state and returns $\sig_{C}$, the $n_C \times n_C$ ``noisy''
encoded state. Finally, $\rho_R$ is the $n_R\times n_R$ dimensional
{\em recovered state}. The objective considered here is to design
$(\Ccal,\Rcal)$ so that the map $\rho_S \to \rho_R$ is as close as
possible to a desired $n_S\times n_S$ unitary $\udes_S$. Hence,
$\rho_R$ has the same dimension as $\rho_S$, that is, $n_R=n_S$. For
emphasis we will replace $\rho_R$ with $\rhoh_S$.

Although it is possible for $\Ecal$ to be non-trace preserving, in the
model considered here, all three quantum operations are each
characterized by a trace-preserving operator-sum-representation (OSR):
\beq[eq:cer osr]
\bea{lll}
\ds
\rho_C=\Ccal(\rho_S)=\sum_c\ C_c \rho_S C_c^\adj,
&\ds
\sum_c\ C_c^\adj C_c = I_{n_S},
&\ds
C_c \in\Cbf^{n_C\times n_S}
\\&&\\
\ds
\sig_C=\Ecal(\rho_C)=\sum_e\ E_e \rho_C E_e^\adj,
&\ds
\sum_e\ E_e^\adj E_e = I_{n_C},
&\ds
E_e \in\Cbf^{n_C\times n_C}
\\&&\\
\ds
\rhoh_S=\Rcal(\sig_C)=\sum_r\ R_r \sig_C R_r^\adj,
&\ds
\sum_r\ R_r^\adj R_r = I_{n_C},
&\ds
R_r \in\Cbf^{n_S\times n_C}
\eea
\eeq
These engender a single trace-preserving quantum operation, $\Scal$,
mapping $\rho_S$ to $\rhoh_S$,
\beq[eq:scal]
\bea{rcl}
\ds
\rhoh_S 
&=& 
\ds
\Scal(\rho_S)
=
\sum_{r,e,c}\
S_{rec}\ \rho_S S_{rec}^\adj
\\&&\\
\ds
S_{rec} 
&=& 
\ds
R_r E_e C_c \in\Cbf^{n_S\times n_S}
\; \Rightarrow\;
\sum_{r,e,c}\ S_{rec}^\adj S_{rec} = I_{n_S} 
\eea
\eeq
Before we describe our design approach we make a few remarks about the
error source and implementation of the encoding and recovery
operations.

\subsection{Implementation}

\begin{figure}[h]
\centering
\epsfig{file=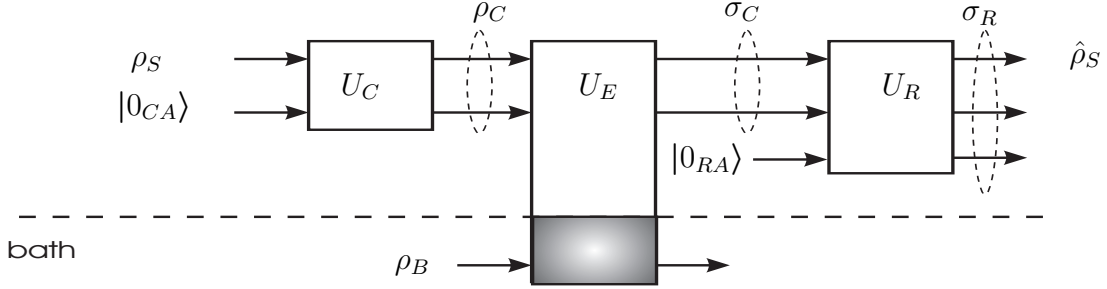,height=1.5in,width=6in}
\caption{System-ancilla-bath representation of standard
encoding-error-recovery model of error correction system.}
\label{fig:urec}
\end{figure}

Any OSR can be equivalently expressed, and consequently physically
implemented, as a unitary with ancilla states \cite[\S
8.23]{NielsenC:00}. An equivalent system-ancilla-bath representation
of the standard error correction model of Figure \ref{fig:rec} is
shown in the block diagram of Figure \ref{fig:urec}.

\psfrag{uc}{$U_C$}
\psfrag{ue}{$U_E$}
\psfrag{ur}{$U_R$}
\psfrag{psica}{$\ket{0_{CA}}$}
\psfrag{rhob}{\!\!$\rho_B$}
\psfrag{rhor}{$\sig_R$}
\psfrag{rhohq}{$\rhoh_S$}
\psfrag{anc}{\hspace{-2ex}Ancilla}

For the encoding operator, $\Ccal$, the {\em encoding ancilla state},
$\ket{0_{CA}}$, has dimension $n_{CA}$, and hence, the resulting
encoded space has dimension $n_C = n_S\ n_{CA}$.  The encoding
operation is determined by $U_C$, the $n_C\times n_C$ unitary encoding
operator which produces the encoded state
$
\sig_C = U_C(\rho_S \otimes \ket{0_{CA}}\bra{0_{CA}} ) U_C^\adj
$
. 

For the error system, $\Ecal$, the ancilla states are engendered by
interaction with the {\em environment}, or the term used here, the
{\em bath}. Here we will ignore complications associated with an
infinite dimensional bath. The error system is thus equivalent to the
$n_E\times n_E$ unitary error operator $U_E$ with uncoupled inputs,
$\rho_{C}$ the encoded state, and $\rho_B$, the $n_B\times n_B$ bath
state.  Thus, $n_E = n_S n_{CA} n_B$.  The noisy encoded state
$\sig_{C}$, is the $n_C\times n_C$ reduced state obtained by tracing
out the bath from the output of $U_E$, that is,
$
\sig_C=\trace_B\ U_E(\rho_C \otimes \rho_B) U_E^\adj
$
. 
 
The recovery system $\Rcal$ has additional ancilla $\ket{0_{RA}}$ of
dimension $n_{RA}$.  $U_R$ is the $n_R\times n_R$ unitary recovery
operator with $n_R=n_S n_{CA} n_{RA}$ and with $\sig_R$ the $n_R\times
n_R$ full output state
$
\sig_R = U_R(\sig_C \otimes \ket{0_{RA}}\bra{0_{RA}} ) U_R^\adj
$
. The $n_S\times n_S$ reduced output state, $\rhoh_S$, is given by the
partial trace over {\em all} the ancillas, the bath having been traced
out in the previous step. Specifically,
$
\rhoh_S = \trace_{A}\ \sig_R
$

\paragraph{Caveat emptor}

\begin{figure}[h]
\centering
\epsfig{file=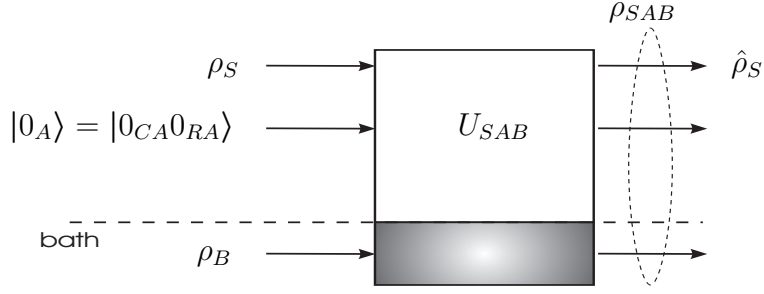,height=1.5in,width=4in}
\caption{System-ancilla-bath representation of error correction
system.}
\label{fig:uqab}
\end{figure}

The ``real'' error correction system is unlikely to be accurately
represented by the system shown in Figure \ref{fig:urec}, but rather
by a full {\em system-ancilla-bath} interaction \cite{AlickiLZ:05}. As
shown in the block diagram in Figure \ref{fig:uqab}, $U_{QAB}$ is the
$n_{QAB}\times n_{QAB}$ unitary system-ancilla-bath operator,
$\ket{0_{CA}0_{RA} }$ is the total ancilla state of dimension $n_{CA}
n_{RA}$ and $\rho_B$ is the bath state. The reduced system output
state, $\rhoh_S$, is obtained from the full output state $\rho_{QAB}$
by tracing simultaneously over all the ancilla and the bath,
$
\rhoh_S = \trace_{AB}\ \rho_{QAB}
$
. At this level of representation, there is no distinction between the
$n_{CA}$ ancilla states used for encoding and the $n_{RA}$ ancilla
states used for recovery. The internal design, however, may be
constructed with such a distinction.

\psfrag{psia}{
\hspace{-15ex}
$\ket{0_A}=\ket{0_{CA}0_{RA} }$
}
\psfrag{uqab}{\!\!$U_{SAB}$}
\psfrag{rhoqab}{$\rho_{SAB}$}

\subsection{Optimal error correction: maximizing fidelity}

As stated, the goal is to make the operation $\rho_S \to \rhoh_S$ be
as close as possible to a desired unitary operation $\udes_S$.
Measures to compare two quantum channels are typically based on
fidelity or distance, \eg, \cite{GilchristLN:04}, \cite{KosutGBR:06}.
Let $\Scal$ denote a trace-preserving quantum channel mapping
$n$-dimensional states to $n$-dimensional states,
\beq[eq:srho] 
\Scal(\rho) = \sum_k S_k \rho S_k^\adj,     
\;\;\;
\sum_k S_k^\adj S_k = I_n
\eeq
The following fidelity inequalities hold:
\beq[eq:fid ineq] 
\fmix \leq \fpur \leq \favg
\eeq
where
\beq[eq:fid def]
\bea{rcl}
\fmix 
&=& 
\min_{\rho} \sum_k | \trace\ S_k \rho |^2
\\&&\\
\fpur 
&=& 
\min_{\ket{\psi}}  \sum_k | \bra{\psi} S_k \ket{\psi} |^2
\\&&\\ 
\favg 
&=& 
\frac{1}{n^2} \sum_k | \trace S_k  |^2
\eea
\eeq
All are in [0,1] and equal to one if and only if $\Scal(\rho)=\rho$.
From \refeq{scal}, $\Scal=\udes_S^\adj \Rcal\Ecal\Ccal$ with OSR
elements $S_k = S_{rec}=\udes_S^\adj R_r E_e C_c \in \Cbf^{n_S\times
n_S}$. Thus $\favg=1$ $\Leftrightarrow$ $\Scal=\mcal{I}_S$
$\Leftrightarrow$ $\Scal = \Rcal\Ecal\Ccal = \udes_S$.

Given $\Scal$, not all these fidelity measures are easy to
calculate. Specifically, $\fmix$ is a convex optimization over all
densities, that is, over all $\rho\in\Cbfnn,\ \rho\geq 0,\ \trace\
\rho=1$, and hence, can be numerically obtained. Calculation of
$\favg$ is direct.  Calculating $\fpur$ is, unfortunately, not a
convex optimization over all pure states $\ket{\psi}$. If, however,
the density associated with $\fmix$ is nearly rank one, then $\fmix
\approx \fpur$.

As a practical matter, when dealing with small channel errors, it does
not matter which fidelity measure is used.
%
%
%
Therefore it is convenient to use $\favg$, as it is already in a form
explicitly dependent only on the OSR elements. In \cite{ReimpellW:05}
$\favg$ was also used as the design measure, but specific convex
optimization algorithms were not proposed. In \cite{YamamotoHT:05} a
similar optimization was proposed using a distance measure to obtain
the recovery given the encoding.

We now focus on the optimization problem,
\beq[eq:favg opt]
\bea{ll}
\mbox{maximize}
&
\favg(\Rcal,\Ccal)
=
\frac{1}{n_S^2}
\sum_{r,e,c}\ \abs{ \trace\ \udes_S^\adj R_r E_e C_c }^2
\\&\\
\mbox{subject to}
&
\sum_r\ R_r^\adj R_r = I_{n_C},
\;\;\; 
\sum_r\ C_c^\adj C_c = I_{n_S}
\eea
\eeq
The optimization variables are the OSR elements $\seq{C_c}$ and
$\seq{R_r}$. As posed this is a difficult optimization problem. The
objective function is not a convex function of either of the design
variables. In addition, the equality constraints are quadratic, and
hence, not convex sets. The problem, however, can be approximated
using {\em convex relaxation}, where each nonconvex constraint is
replaced with a less restrictive convex constraint
\cite{BoydV:04}. This finally results in a bi-convex optimization
problem in the encoding and recovery operator elements which can be
iterated to yield a local optimum.  As we show next, iterating between
the two problems is guaranteed to increase fidelity of each of the
relaxed problems.  When the iterations converge, all that can be said
is that a local solution has been found.

\section{Optimal error correction via bi-convex relaxation}

\subsection{Process matrix problem formulation}

Following the procedure used in quantum process tomography \cite[\S
8.4.2]{NielsenC:00}, \cite{KosutWR:04} we expand each of the OSR
elements $R_r\in\Cbfsc$ and $C_c\in\Cbfcs$ in a set of basis matrices,
respectively, for $\Cbfsc$ and $\Cbfcs$, that is,
\beq[eq:rcb]
\bea{rcl}
R_r &=& \sum_i\ x_{ri}\ B_{Ri}
\\
C_c &=& \sum_i\ x_{ci}\ B_{Ci}
\eea
\eeq
where $ \set{B_{Ri}\in\Cbfsc,\ B_{Ci}\in\Cbfcs}{i=1,\ldots,n_Sn_C} $
and the $\seq{x_{ri}}$ and $\seq{x_{ci}}$ are complex scalars. 
Problem \refeq{favg opt} can then be equivalently expressed as,
\beq[eq:favg opt xrc]
\bea{ll}
\mbox{maximize}
&
\favg(\Rcal,\Ccal) \equiv \favg(X_R,X_C) 
=
\sum_{ijk\ell}\ (X_R)_{ij}\ (X_C)_{k\ell}\ F_{ijk\ell}
\\&\\
\mbox{subject to}
&
\sum_{ij}\ (X_R)_{ij}\ B_{Ri}^\adj B_{Rj} = I_{n_C}
\\&\\
&
\sum_{k\ell}\ (X_C)_{k\ell}\ B_{Ck}^\adj B_{C\ell} = I_{n_S}
\\&\\
&
(X_R)_{ij} = \sum_r\ x_{ri} x_{rj}^*
\\&\\
&
(X_C)_{k\ell} = \sum_c\ x_{ck} x_{c\ell}^*
\\&\\
&
F_{ijk\ell}
= 
\sum_{e}\
(\trace\ L_S^\adj B_{Ri} E_e B_{Ck})
(\trace\ L_S^\adj B_{Rj} E_e B_{C\ell})^*/n_S^2
\eea
\eeq
The optimization variables are the {\em process matrices} $X_R,\ X_C
\in\Cbfscsc$ and the scalars $\seq{x_{ri}}$ and $\seq{x_{ci}}$. The
problem data which describes the desired unitary and error system is
contained in the $\seq{ F_{ijk\ell} }$.  The equality constraints
$(X_R)_{ij} = \sum_r\ x_{ri} x_{rj}^*$ and $(X_C)_{k\ell} = \sum_c\
x_{ck} x_{c\ell}^*$ are both quadratic, exposing again that this is
not a convex optimization problem. We do not explore the possible
simplifications that can occur in these expressions if the basis
matrices are chosen prudently, \eg, $\trace\ B_i^\adj B_j =
\del_{ij}$.

\subsection{Design of $\Rcal$ given $\Ccal$ and $\Ecal$}

In this section and in the remainder of the paper we set the desired
logical operation to identity, \ie, $\udes_S=I_S$; just error
correction not correction and computation. This is without loss of
generality as a desired logical operation can be added everywhere.

Suppose the encoding $\Ccal$ is given (and $\udes_S=I_S$). Then
optimizing only over $\Rcal$ in \refeq{favg opt xrc} can be
equivalently expressed as,
\beq[eq:favg opt rcal]
\bea{ll}
\mbox{maximize}
&
\favg(\Rcal,\Ccal) \equiv \favg(X_R,\Ccal) 
=
\trace\ X_R W_R(\Ecal,\Ccal)
\\&\\
\mbox{subject to}
&
\sum_{i,j}\ (X_R)_{ij}\ B_{Ri}^\adj B_{Rj} = I_{n_C}
\\&\\
&
(X_R)_{ij} = \sum_r\ x_{ri} x_{rj}^*
\\&\\
&
(W_R(\Ecal,\Ccal))_{ij} 
= 
\sum_{c,k,\ell}\ x_{ck}x_{c\ell}^*\ F_{ijk\ell}
=
\sum_{e,c}\
(\trace\ B_{Ri} E_e C_c)(\trace\ B_{Rj} E_e C_c)^*/n_S^2
\eea
\eeq
The optimization variables are the matrix $X_R\in\Cbfscsc$ and the
scalars $\seq{x_{ri}}$. The problem data is contained in the positive
semidefinite matrix $W_R(\Ecal,\Ccal)\in\Cbfscsc$. The objective
function is now linear in $X_R$, which is of course a convex
function. However, each of the equality constraints, $(X_R)_{ij} =
\sum_r\ x_{ri} x_{rj}^*$ is quadratic, and thus does not form a convex
set. This set of quadratic equality constraints can be relaxed to the
matrix inequality constraint, $X_R\geq 0$, that is, $X_R$ is positive
semidefinite, a convex set in the elements of $X_R$. A convex
relaxation of \refeq{favg opt rcal} is then,
\beq[eq:favg rel rcal]
\bea{ll}
\mbox{maximize}
&
\trace\ X_R W_R(\Ecal,\Ccal)
\\&\\
\mbox{subject to}
&
X_R \geq 0,
\;\;\;
\sum_{i,j}\ (X_R)_{ij}\ B_{Ri}^\adj B_{Rj} = I_{n_C}
\eea
\eeq
This class of convex optimization problems is referred to as an SDP,
for {\em semidefinite program} \cite{BoydV:04}.\footnote{
A standard SDP is to minimize a linear objective function subject to
convex inequalities and linear equalities. The objective function in
\refeq{favg rel rcal} is the maximization of a linear function which
is equivalent to the minimization of its' negative, and hence, is a
linear objective function.
} For a given encoding $\Ccal$, the optimal solution to the relaxed
problem \refeq{favg rel rcal}, $X_R^\rlxopt$, provides an upper bound
on the average fidelity objective in \refeq{favg opt} or \refeq{favg
opt xrc}. From the fidelity inequalities \refeq{fid ineq}, we can
derive a lower bound. Specifically, the (unknown, possibly unknowable)
solution to the original problem \refeq{favg opt}, is bounded as
follows:
\beq[eq:favg rcal bnds]
\fmix(\Rcal^\rlxopt,\Ccal)
\leq
\max_{\Rcal}\ \fpur(\Rcal,\Ccal)
\leq
\favg(\Rcal^\rlxopt,\Ccal)
=
\trace\ X_R^\rlxopt W_RS(\Ecal,\Ccal)
\eeq
where $\Rcal^\rlxopt$ is the OSR with elements $\seq{R_r^\rlxopt}$
obtained from $X_R^\rlxopt$ via the singular value decomposition,
\beq[eq:xr2osr]
X_R^\rlxopt = VSV^\adj
\;\Rightarrow\;
R_r^\rlxopt
=
\sqrt{s_r}\ \sum_{i=1}^{n_S n_C}\ V_{ir} B_{Ri},
\;\;
r=1,\ldots,n_S n_C
\eeq
where $V\in\Cbfscsc$ is unitary and $S=\diag(s_1\ \cdots\ s_{n_S
n_C})$ with singular values in decreasing order, $s_1 \geq s_2 \geq\
\cdots\ \geq s_{n_S n_C} \geq 0 $.

\subsection{Design of $\Ccal$ given $\Rcal$ and $\Ecal$}

Repeating the previous steps, optimizing only over $\Ccal$ in
\refeq{favg opt xrc} can be equivalently expressed as,
\beq[eq:favg opt ccal]
\bea{ll}
\mbox{maximize}
&
\favg(\Rcal,\Ccal) \equiv \favg(X_C,\Rcal) 
=
\trace\ X_C W_C(\Ecal,\Rcal)
\\&\\
\mbox{subject to}
&
\sum_{k,\ell}\ (X_C)_{k\ell}\ B_{Ck}^\adj B_{C\ell} = I_{n_S}
\\&\\
&
(X_C)_{k\ell} = \sum_c\ x_{ck} x_{c\ell}^*
\\&\\
&
(W_C(\Ecal,\Rcal))_{k\ell} 
=\sum_{r,i,j}\
x_{ri}x_{rj}^*\ F_{ijk\ell}
= \sum_{e,r}\
(\trace\ B_{Ck} R_r E_e)(\trace\ B_{C\ell} R_r E_e)^*/n_S^2
\eea
\eeq
The optimization variables are the matrix $X_C\in\Cbfscsc$ and the
scalars $\seq{x_{ci}}$ with all the problem data contained in the
symmetric positive semidefinite matrix $W_C(\Ecal,\Rcal)\in\Cbfscsc$. In
this case, however, the basis matrices, $\seq{B_{Ci}}$ are $n_C\times
n_S$. Repeating the previous procedure of relaxing the quadratic
equality constrain to $X_C \geq 0$, we obtain the convex relaxation of
\refeq{favg opt xrc} as the SDP,
\beq[eq:favg rel ccal]
\bea{ll}
\mbox{maximize}
&
\trace\ X_C W_C(\Ecal,\Rcal)
\\&\\
\mbox{subject to}
&
X_C \geq 0,
\;\;\;
\sum_{i,j}\ (X_C)_{ij}\ B_{Ci}^\adj B_{Cj} = I_{n_S}
\eea
\eeq
Analogously to \refeq{xr2osr}, for a given recovery $\Rcal$, the
(unknown, possibly unknowable) solution to the original problem
\refeq{favg opt}, is bounded as follows:
\beq[eq:favg ccal bnds]
\fmix(\Rcal,\Ccal^\rlxopt)
\leq
\max_{\Ccal}\ \fpur(\Rcal,\Ccal)
\leq
\favg(\Rcal,\Ccal^\rlxopt)
=
\trace\ X_C^\rlxopt W_C(\Ecal,\Rcal)
\eeq
where $\Ccal^\rlxopt$ is the OSR with elements $\seq{C_c^\rlxopt}$
obtained from $X_C^\rlxopt$ via the singular value decomposition,
\beq[eq:xc2osr]
X_C^\rlxopt = VSV^\adj
\;\Rightarrow\;
C_c^\rlxopt
=
\sqrt{s_c}\ \sum_{i=1}^{n_S n_C}\ V_{ic} B_{Ci},
\;\;
c=1,\ldots,n_S n_C
\eeq
where $V\in\Cbfscsc$ is unitary and $S=\diag(s_1\ \cdots\ s_{n_S
n_C})$ with singular values in decreasing order, $s_1 \geq s_2 \geq\
\cdots\ \geq s_{n_S n_C} \geq 0 $.

\subsection{Iterative bi-convex algorithm}

Proceeding analogously as in \cite{ReimpellW:05}, the two separate
optimizations for $\Ccal$ and $\Rcal$ can be combined into the
following iteration.

\blist
\item {\bf initialize} encoding $\hat{\Ccal}$
and stopping level $\epsilon$
\item {\bf repeat}
	\ben
	\item {\bf optimize recovery}
		\ben
		\item compute $X_R^\star$ as solution to:
\[
\bea{ll}
\mbox{maximize}
&
\trace\ X_R W_R(\Ecal,\hat{\Ccal})
\\&\\
\mbox{subject to}
&
X_R \geq 0,
\;\;\;
\sum_{i,j}\ (X_R)_{ij}\ B_{Ri}^\adj B_{Rj} = I_{n_C}
\eea
\]
		\item use \refeq{xr2osr} to compute $\Rcal^\star$ from
		$X_R^\star$
		\een

	\item {\bf optimize encoding}
		\ben	
		\item compute $X_C^\star$ as solution to:
\[
\bea{ll}
\mbox{maximize}
&
\trace\ X_C W_C(\Ecal,\Rcal^\star)
\\&\\
\mbox{subject to}
&
X_R \geq 0,
\;\;\;
\sum_{i,j}\ (X_C)_{ij}\ B_{Ci}^\adj B_{Cj} = I_{n_S}
\eea
\]
		\item use \refeq{xc2osr} to compute $\Ccal^\star$ from
		$X_C^\star$
		\een

	\item {\bf compute change in fidelity}
\[
\Del\favg  = \favg(\Rcal^\star,\Ccal^\star)-\favg(\Rcal^\star,\hat{\Ccal})
\]
	\item {\bf reset}
\[
\hat{\Ccal} = \Ccal^\star
\]
	\een
\item {\bf until} 
\[
\Del\favg < \epsilon
\]
\elist
The algorithm returns $(\Rcal^\star,\Ccal^\star)$. The optimization in
each of the steps is a convex optimization and hence fidelity will
increase in each step, thereby converging to a local solution of the
joint relaxed problem. This solution is not necessarily a local
solution to the original problem \refeq{favg opt} or \refeq{favg opt
xrc}. However, the upper and lower bounds obtained will apply.  The
optimization steps can be reversed by starting with an initial
recovery and then starting the iteration by optimizing over encoding.

\subsection{Decoherence resistant encoding}

If the sole purpose of encoding is to sustain the information state
$\rho_S$, then the desired operation is the identity ($\udes_S=I_S$)
and the recovery operation in Figure \ref{fig:rec} is simply the
partial trace over the encoding ancilla, that is,
\beq[eq:rcal partr]
\rhoh_S
=
\Rcal(\sig_C) 
= 
\trace_{CA}\ \sig_C  
=
\left[
\bea{ccc}
\trace\ (\sig_C)_{[1,1]} & \cdots & \trace\ (\sig_C)_{[1,n_S]}
\\
\vdots & \vdots & \vdots
\\
\trace\ (\sig_C)_{[n_S,1]} & \cdots & \trace\ (\sig_C)_{[n_S,n_S]}
\eea
\right]
\eeq
where the $(\sig_C)_{[i,j]}$ are the $n_S^2$ sub-block matrices of
$\sig_C$, each being $n_{CA}\times n_{CA}$. Hence, the OSR elements of
$\Rcal$ are given by
\beq[eq:rosr partr]
(R_r)_{ij}
=
\left\{
\bea{ll}
1 & j=(i-1)n_{CA}+r
\\
0 & \mbox{else}
\eea
\right\},
\;
r=1,\ldots,n_{CA},\
j=1,\ldots,n_S
\eeq
For a given error $\Ecal$, finding an optimal encoding by solving
\refeq{favg rel ccal} is equivalent to finding a {\em
decoherence-resistant-subspace}. If there is perfect recovery, then we
have found a {\em decoherence-free-subspace} \cite{LidarCW:98}.  In
\cite{ZanardiL:04}, this problem was considered using $\fpur$, the
pure state fidelity.

\subsection{Robust error correction}
\label{sec:robust}

The bi-convex optimization can be extended to the case where the error
system is one of a number of possible error systems, that is,
\beq[eq:ecal alf]
\Ecal \in \set{\Ecal_\alf}{\alf=1,\ldots,\ell}
\eeq
where each $\Ecal_\alf$ has OSR elements $\seq{E_{\alf e}}$. The
worst-case fidelity design problem, by analogy with \refeq{favg opt
xrc}, is then:
\beq[eq:robust]
\bea{ll}
\mbox{maximize}
&
\min_\alf\ \favg(\Rcal,\Ecal_\alf,\Ccal)
=
\sum_{ijk\ell}\ (X_R)_{ij}\ (X_C)_{k\ell}\ F_{\alf ijk\ell}
\\&\\
\mbox{subject to}
&
\mbox{$X_R,\ X_C$ constrained as in \refeq{favg opt xrc}}
\\&\\
&
F_{\alf ijk\ell}
= 
\sum_{e}\
(\trace\ L_S^\adj B_{Ri} E_{\alf e} B_{Ck})
(\trace\ L_S^\adj B_{Rj} E_{\alf e} B_{C\ell})^*/n_S^2
\eea
\eeq
Iterating as before between $\Rcal$ and $\Ccal$ results again in
separate convex optimization problems, each of which is an SDP.
Specifically, for a given encoding $\Ccal$, a robust recovery is
obtained from,
\beq[eq:favg rob r]
\bea{ll}
\mbox{maximize}
&
\min_\alf\ \trace\ X_R W_R(\Ecal_\alf,\Ccal)
\\&\\
\mbox{subject to}
&
X_C \geq 0,
\;\;\;
\sum_{i,j}\ (X_C)_{ij}\ B_{Ci}^\adj B_{Cj} = I_{n_S}
\eea
\eeq
Similarly, for a given recovery $\Rcal$, a robust encoding is obtained
from,
\beq[eq:favg rob c]
\bea{ll}
\mbox{maximize}
&
\min_\alf\ \trace\ X_C W_C(\Ecal_\alf,\Rcal)
\\&\\
\mbox{subject to}
&
X_R \geq 0,
\;\;\;
\sum_{i,j}\ (X_R)_{ij}\ B_{Ri}^\adj B_{Rj} = I_{n_C}
\eea
\eeq
%

\section{Computing the solution: Lagrange Duality}

The main difficulty with embedding the OSR elements into either $X_R$
or $X_C$ is {\em scaling with qubits}. Specifically, the number of
design parameters needed to determine either $X_C$ or $X_R$ scales
exponentially with the number of qubits. Although exponential scaling
at the moment seems unavoidable, we show in this section that solving
the dual SDPs associated with either \refeq{favg rel rcal} or
\refeq{favg rel ccal} requires many fewer parameters, and thus
engenders a reduced computational burden.

The convex optimization problems \refeq{favg rel rcal} and \refeq{favg
rel ccal} are both SDPs of the form,
\beq[eq:primal sdp]
\bea{ll}
\mbox{maximize}
&
\trace\ XW
\\
\mbox{subject to}
&
X \geq 0,
\;\;
\sum_{ij} X_{ij} B_i^\adj B_j = I_m
\eea
\eeq
with optimization variable $X = X^\dag \in \Cbfnn$, $n=rm$ for some
integer $r$, and with each basis matrix $B_i\in\Cbf^{r\times m}$.  We
will refer to this as the {\em primal problem}.  From \cite[\S
11.8.3]{BoydV:04}, for the standard SDP: $\mbox{minimize}\ c^Tx,\
\mbox{subject to}\ F_0+\sum_i\ x_i F_i \geq 0$ with $x\in\Rbf^p$ and
$F_i=F_i^T\in\Rbf^{q\times q}$, the computational complexity using a
primal-dual algorithm is $\max\seq{pq^3,p^2q^2,p^3}$ flops (floating
point operations) per iteration step where typically 10-100 steps are
required in the algorithm.  Accounting for the linear (matrix)
equality constraint and the Hermiticity of $X$, the number of real
optimization variables in \refeq{primal sdp} is
$p=n^2-m^2=(r^2-1)m$. The dimension of the linear matrix inequality is
$q=n=rm$. This gives the computational complexity as
$p^2q^2=r^2(r^2-1)^2m^6$ flops per iteration.

Solving \refeq{favg rel rcal} for $X_R$, gives $n=n_Sn_C,\ r=n_S,\
m=n_C=n_Sn_{CA}$ for $n_S^8(n_S^2-1)^2n_{CA}^6$ flops.  Solving
\refeq{favg rel ccal} for $X_C$, gives $n=n_Sn_C,\ r=n_C,\ m=n_S$ for
$n_S^8(n_S^2n_{CA}^2-1)^2n_{CA}^2$ flops. Exponential growth in
computation occurs becasue each of these dimensions are exponential in
the number of qubits, \ie, , $n_S=2^{q_S},\ n_C=2^{q_S+q_{CA}}$, and
so on.


The computational burden can be somewhat alleviated by appealing to
{\em Lagrange Duality Theory} \cite[Ch.5]{BoydV:04} which provides a
means for establishing a lower bound on the optimal objective value,
establishing conditions of optimality, and providing, in some cases,
and this case in particular, a more efficient means to numerically
solve the original problem.  In Appendix \ref{sec:dual} we show that
the {\em dual problem} associated with the {\em primal problem}
\refeq{primal sdp} is,
\beq[eq:dual sdp]
\bea{ll}
\mbox{minimize}
&
\trace\ Y
\\
\mbox{subject to}
&
K(Y) - W \geq 0,
\;\;
K_{ij}(Y) = \trace\ B_j^\adj B_i Y
\eea
\eeq
with optimization variable $Y=Y^\dag\in\Cbfmm$. The number of (real)
optimization variables for the dual problem is then at most $m^2$. The
dual problem is also an SDP and from the previous formula therefore
requires $r^2m^6$ flops per iteration, a reduction in flops per
iteration from the primal by a factor of $(r^2-1)^2$.  We show in
Appendix \ref{sec:dual} that if $(X^\opt, Y^\opt)$ solve the primal
and dual problems respectively, then:
\beq[eq:optcon]
\bea{rcl}
\trace\ X^\opt W &=& \trace\ Y^\opt
\\
(K(Y^\opt)-W)X^\opt &=& 0
\eea
\eeq
The second equation above together with the linear equality constraint
in \refeq{primal sdp} can be used to obtain the primal solution
$X^\opt$ from the dual solution $Y^\opt$. That is, solve for $X^\opt$
from the set of {\em linear} equations,
\beq[eq:d2p]
\bea{rcl}
(K(Y^\opt)-W)X^\opt &=& 0
\\
\sum_{ij} X^\opt_{ij} B_i^\adj B_j &=& I_m
\eea
\eeq
Solving this type of linear set of equations is an {\em eigenvalue
problem} and thus requires on the order of no more than $n^2$ flops
\cite{GolubV:83}. Thus the dual takes $r^2m^6$ flops per iteration
plus $r^2m^2$ flops one time to convert from dual to primal. This is
in comparison to the much larger $r^2(r^2-1)^2m^6$ flops per iteration
for the primal alone. Neglecting the dual to primal conversion, the
speed-up in flops per iteration to calculate $X_R$ is approximately
$(n_S^2-1)^2,\ n_s=2^{q_S}$ and for $X_C$ it is $(n_C^2-1)^2,\
n_C=2^{q_S+q_{CA}}$.

\section{Example}


In this illustrative example, the goal is to preserve a single
information qubit using a single ancilla qubit. Thus, the desired
logical gate is the identity, that is, $L_S=I_2$, with $n_S = n_{CA} =
2$, and hence, $n_C =4$. We made two error systems, $\Ecal_a$ and
$\Ecal_b$, by randomly selecting the unitary bath representation as
shown in Figure \ref{fig:urec} as follows: Each error system has a
single qubit bath state, $\ket{0}_B$, thus $n_B=2$.  The Hamiltonian
for each system, $H_E=H_E^\adj\in\Cbf^{n_E\times n_E},\ n_E=n_Cn_B=8$,
was chosen randomly and then adjusted to have the magnitude (maximum
singular value) $\norm{H_E}=\del_E =0.75$. Then the unitary
representing the error system was set to $U_E=\exp(-i H_E)$ and from
this the corresponding OSR $\Ecal$ was computed.  The OSR elements to
three decimal points for the two random systems are as follows.
%
\begin{small}
\[
\Ecal_a\ 
\left\{ 
\bea{rcl} 
E_{a1} &=& 
\left[ 
\bea{cccc} 
0.9-0.049i & 0.193+0.194i & -0.161+0.039i & -0.135+0.156i\\ 
-0.159+0.148i & 0.887-0.046i & 0.148-0.025i & -0.168-0.081i\\ 
0.167+0.061i & -0.07+0.004i & 0.905+0.161i & 0.16+0.125i\\ 
0.124+0.137i & 0.167-0.155i & -0.203+0.118i & 0.844-0.26i\\
\eea
\right]
\\&&\\
E_{a2} &=&
\left[
\bea{cccc}
0.053-0.063i & -0.034+0.082i & 0.148-0.085i & 0.13-0.076i\\
-0.168-0.01i & 0.141+0.073i & 0.008+0.091i & -0.074-0.024i\\ 
0.119+0.053i & 0.207-0.02i & 0.043+0.01i & -0.063-0.21i\\ 
0.123+0.066i & 0.027+0.008i & 0.07-0.058i & 0.098-0.11i\\
\eea
\right]
\eea
\right.
\]
\[
\Ecal_b\ 
\left\{ 
\bea{rcl} 
E_{b1} &=& 
\left[ 
\bea{cccc} 
0.943+0.018i & -0.14-0.024i & 0.076-0.081i & 0.04-0.163i\\ 
0.107+0.062i & 0.876+0.068i & -0.06-0.021i & -0.127+0.06i\\ 
-0.025-0.042i & 0.122+0.073i & 0.889-0.035i & 0.043-0.078i\\ 
-0.017-0.094i & 0.095+0.035i & -0.032-0.089i & 0.88+0.113i\\
\eea
\right]
\\&&\\
E_{b2} &=&
\left[
\bea{cccc}
0.07i & -0.2-0.082i & 0.028-0.083i & 0.179+0.206i\\ 
-0.003-0.147i & 0.138-0.155i & 0.202+0.306i & 0.045-0.134i\\ 
0.049+0.084i & -0.149+0.217i & 0.143-0.04i & 0.024+0.174i\\ 
-0.191+0.095i & -0.081-0.097i & 0.007-0.127i & 0.035-0.167i\\
\eea
\right]
\eea
\right.
\]
\end{small}
Neither of these error systems is of the standard type, \eg, there is
no independent channel structure.  The choice of $\del_E=0.75$ is
perhaps extreme, but is motivated here by our desire to demonstrate
that the optimization procedure can handle errors that are beyond the
range of classically-inspired quantum error correction.  For this
particular set of error systems, we do not know if there exists an
encoding/recovery pair limited to using a single encoding ancilla
state which can bring perfect correction.  This also motivates the
search for the still elusive black-box error correction discussed in
the introduction.

For each of the error systems we ran the bi-convex iteration 100 times
starting with the initial recovery operator $\Rcal_0$ given by the
partial trace operation \refeq{rosr partr}. Denote
$(\Rcal_{a1},\Ccal_{a1})$ and $(\Rcal_{a100},\Ccal_{a100})$ as the 1st
and 100th iteration pairs optimized for $\Ecal_a$, and similarly
$(\Rcal_{b1},\Ccal_{b1})$ and $(\Rcal_{b100},\Ccal_{b100})$ as the 1st
and 100th iteration pairs optimized for $\Ecal_b$. Table \ref{tab:favg
ab} shows the average fidelities $\favg(\Rcal,\Ecal,\Ccal)$ for some
of the possible combinations.

\begin{table}[t]
\centering
\renewcommand{\arraystretch}{2}
\begin{tabular}{|c|c||c|c|}
\hline
Type
&
$\Rcal,\ \Ccal$ & $\Ecal_a$ & $\Ecal_b$ 
\\
\hline\hline
Optimal encoding, no recovery
&
$\Rcal_0,\ \Ccal_{a1}$ & 0.9686 & 0.7631
\\
\hline
Optimal encoding \& recovery: 1 iteration
&
$\Rcal_{a1},\ \Ccal_{a1}$ & 0.9719 & 0.7805
\\
\hline
Optimal encoding \& recovery: 100 iterations
&
$\Rcal_{a100},\ \Ccal_{a100}$ & 0.9997 & 0.6261
\\
\hline\hline
Optimal encoding, no recovery
&
$\Rcal_0,\ \Ccal_{b1}$ & 0.7445 & 0.9091
\\
\hline
Optimal encoding \& recovery: 1 iteration
&
$\Rcal_{b1},\ \Ccal_{b1}$ & 0.7843 & 0.9441
\\
\hline
Optimal encoding \& recovery: 100 iterations
&
$\Rcal_{b100},\ \Ccal_{b100}$ & 0.7412 & 0.9997
\\
\hline
\end{tabular}
\caption{Average fidelities}
\label{tab:favg ab}
\end{table}

As Table \ref{tab:favg ab} clearly shows, fidelity tuned for a
specific error, either $\Ecal_a$ or $\Ecal_b$ in this example,
saturated to the levels shown (0.9997) in about 100 iterations.
However, neither of the optimized codes are {\em robust}. Each does
very poorly when the error is different then what was expected.  By
raising the number of ancilla it is of course possible to make the
system robust. This, however, introduces considerable complexity. What
the table suggests is that an alternate route is to tune for maximal
fidelity, say, in a particular module. This of course can only be done
on the actual system.

For each of the optimizations, the process matrices $X_C,\
X_R\in\Cbf^{8\times 8}$, associated respectively with each $\Ccal$ and
$\Rcal$ were of reduced rank. For all the optimized $\Ccal$, each
process matrix $X_C$ was found to have a single dominant singular
value, and hence, there is a single dominant $4\times 2$ OSR element
which characterizes $\Ccal$.  For the optimized $\Rcal$, each $X_R$
was found to have two dominant singular values, and hence, there are
two dominant $2\times 4$ OSR elements which characterize $\Rcal$.  For
example, the recovery/encoding pair $(\Rcal_{a100},\Ccal_{a100})$ has
the OSR elements:
\begin{small}
\[
\bea{rcl}
C_1
&=&
\left[
\bea{cc}
-0.629 & 0.189-0.332i\\ 
0.455+0.378i & 0.207+0.24i\\ 
0.42+0.063i & -0.425-0.358i\\ 
0.13+0.233i & 0.626+0.226i\\
\eea
\right]
\\&&\\
R_1
&=&
\left[
\bea{cccc}
-0.707 & 0.532-0.342i & 0.194-0.175i & 0.103-0.138i\\ 
0.134+0.087i & 0.009-0.166i & -0.103+0.404i & 0.833-0.276i\\
\eea
\right]
\\&&\\
R_2
&=&
\left[
\bea{cccc}
-0.603 & -0.528+0.461i & -0.262-0.131i & 0.172-0.163i\\ 
0.313-0.103i & 0.259+0.104i & -0.374-0.728i & 0.174-0.333i\\
\eea
\right]
\eea
\]
\end{small}
It is not obvious that these correspond to any of the standard codes.
However, by construction, $C_1^\adj C_1=I_2$ and $\sum_{i=1}^2
R_i^\adj R_i=I_4$. Referring to Figure \ref{fig:urec}, we can
construct the encoding and recovery unitaries as,
\[
U_C
=
\left[ C_1\ C_2 \right],
\;\;
U_R = \left[\bea{c} R_1 \\ R_2 \eea\right]
\]
where $C_2\in\Cbf^{2\times 2}$ is arbitrary as long as $U_C$ is
unitary, or equivalently, $C_1^\adj C_2=0$ and $C_2^\adj
C_2=I_2$. Observe that $U_R$ is already a $4\times 4$ unitary.

\begin{figure}[t]
\centering
\epsfig{file=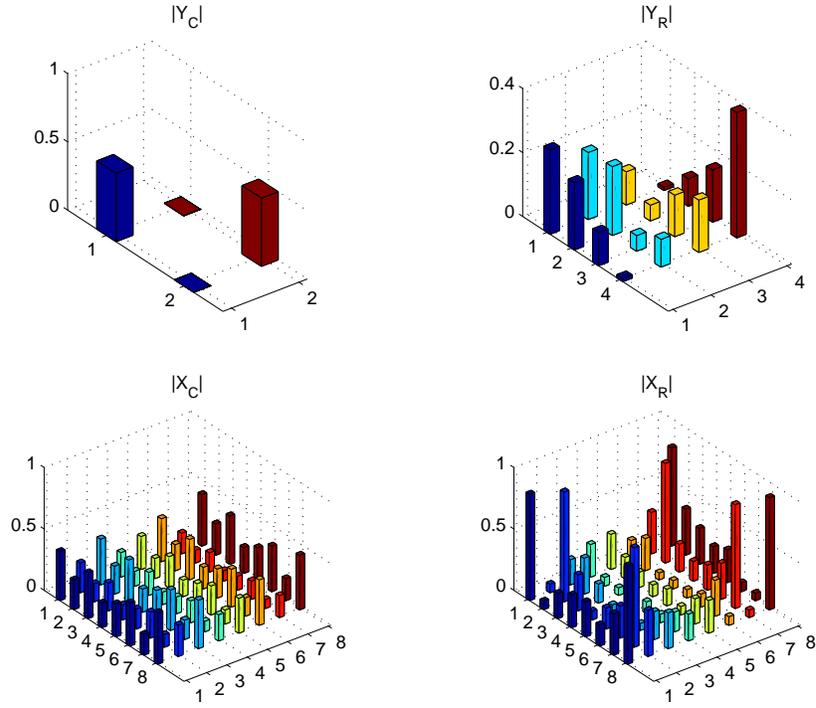,height=3.75in}
\caption{Magnitudes of primal-dual pairs $(X_C,Y_C)$ and $(X_R,Y_R)$
corresponding to $(\Rcal_{a100},\Ccal_{a100})$ optimized for
$\Ecal_a$.}
\label{fig:xya}
\end{figure}

\begin{figure}[b]
\centering
\epsfig{file=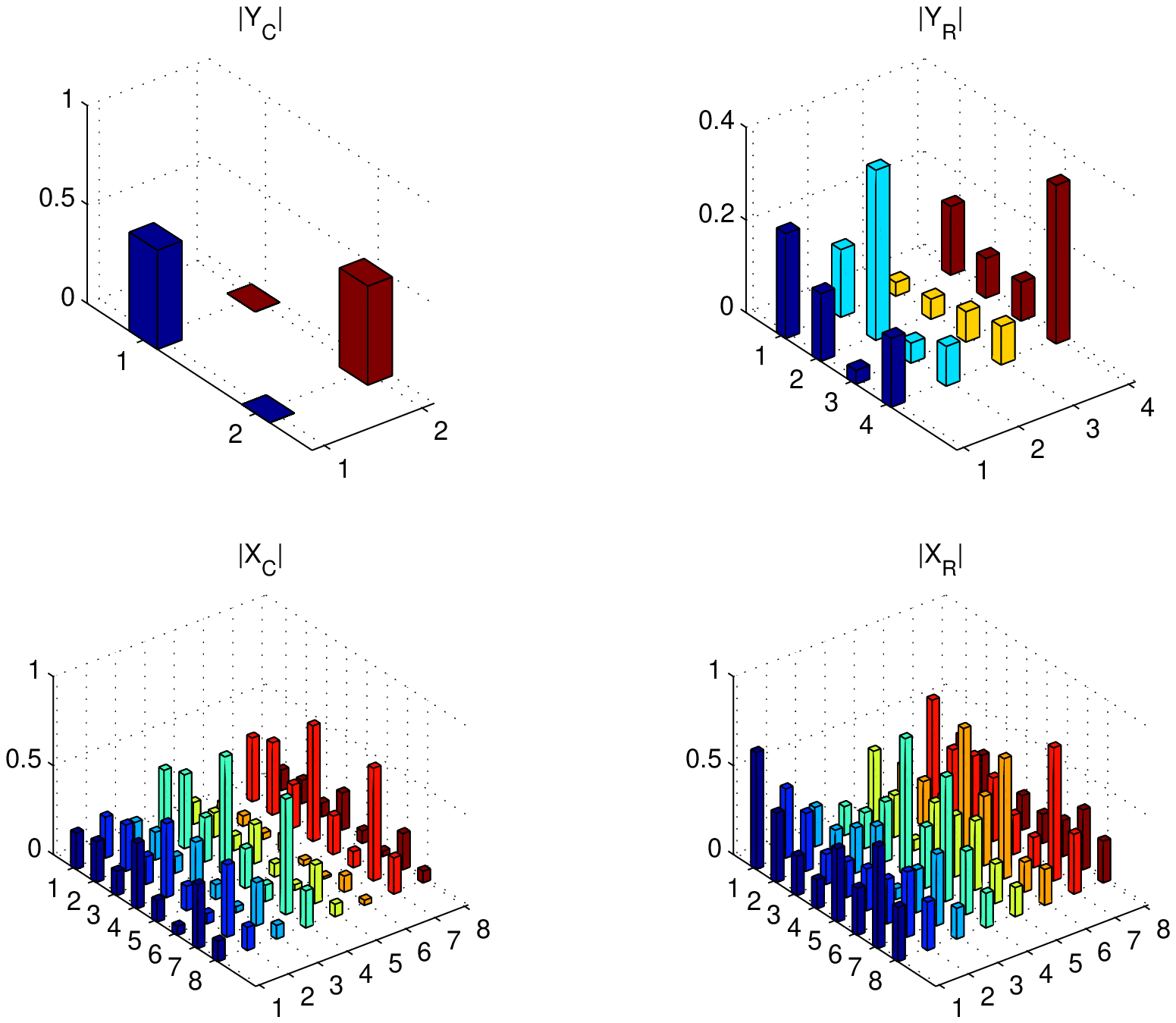,height=3.75in}
\caption{Magnitudes of primal-dual pairs $(X_C,Y_C)$ and $(X_R,Y_R)$
corresponding to $(\Rcal_{b100},\Ccal_{b100})$ optimized for
$\Ecal_b$.}
\label{fig:xyb}
\end{figure}

Bar plots of the magnitude of the elements in the primal-dual pairs
$(X_C,Y_C)$ and $(X_R,Y_R)$ corresponding to
$(\Rcal_{a100},\Ccal_{a100})$ and $(\Rcal_{b100},\Ccal_{b100})$ are
shown in figures \ref{fig:xya} and \ref{fig:xyb}, respectively. From
many of such similar plots we have observed some common structure
which may be used to reduce the computational burden. 

We also computed a {\em robust} encoding and recovery for the error
set $\seq{\Ecal_a,\ \Ecal_b}$ by iterating between \refeq{favg rob r}
and \refeq{favg rob c}. The resultant average fidelities are in Table
\ref{tab:favg rob ab}.

\begin{table}[t]
\centering
\renewcommand{\arraystretch}{2}
\begin{tabular}{|c||c||c|c|}
\hline
Type
&
$\Rcal,\ \Ccal$ & $\Ecal_a$ & $\Ecal_b$ 
\\
\hline\hline
Robust encoding, no recovery
&
$\Rcal_0,\ \Ccal_{ab1}$ & 0.8840 & 0.8840
\\
\hline
Robust encoding \& recovery: 1 iteration
&
$\Rcal_{ab1},\ \Ccal_{ab1}$ & 0.9284 & 0.9284
\\
\hline
Robust encoding \& recovery: 100 iterations
&
$\Rcal_{ab100},\ \Ccal_{ab20}$ & 0.9576 & 0.9576
\\
\hline
\end{tabular}
\caption{Average robust fidelities}
\label{tab:favg rob ab}
\end{table}

Comparing Tables \ref{tab:favg ab} and \ref{tab:favg rob ab} clearly
shows that a robust design is possible although at a cost of
performance. Also in this case after 100 iterations the robust
fidelity did not increase. In addition, the rank of the process
matrices $X_C$ and $X_R$ remained as before at 1 and 2, respectively,
and the resulting OSR elements do not appear standard.

\section{Conclusions}

We have shown that the design of a quantum error correction system can
be cast as a bi-convex iteration between encoding and recovery, each
being a semidefinite program (SDP). We have also shown that the dual
optimization, also an SDP, is of lower complexity and thus requires
less computational effort. The SDP formalism also allows for a robust
design by enumerating constraints associated with different error
models.  We illustrated the approach with an example where the error
system does not assume independent channels.

\paragraph{Note added}

While this work was finalized for submission we became aware of the
closely related \cite{FletcherSW:06} and the subsequent commentary
\cite{ReimpellWA:06}.

\bc\subsection*{Acknowledgements}\ec

This work has been funded under the DARPA QuIST Program (Quantum
Information Science \& Technology) and (to D. A. L.) NSF CCF-0523675
and ARO (Quantum Algorithms, W911NF-05-1-0440).  We would like to
thank Ian Walmsley and Dan Browne (Oxford), Constantin Brif, Matthew
Grace, and Hersch Rabitz (Princeton), and Alireza Shabani (USC) for
many fruitful discussions.

\newpage

\clearpage


\bibliographystyle{alpha}
\bibliography{D:/robert/tex/rlk}

\begin{appendix}

\section{Dual problem}
\label{sec:dual}

We apply Lagrange Duality Theory \cite[Ch.5]{BoydV:04}.
Write the primal problem \refeq{primal sdp} as a minimization,
\beq[eq:prim]
\bea{ll}
\mbox{minimize}
&
-\trace\ XW
\\
\mbox{subject to}
&
X \geq 0,
\;\;
\sum_{ij} X_{ij} C_{ij} = I_m
\eea
\eeq
with optimization variable $X = X^\dag \in \Cbfnn$. The {\em
Lagrangian} for \refeq{prim} is,
\beq[eq:lag]
\bea{rcl}
L(X,Z,Y) 
&=& 
-\trace\ XW =\trace\ XZ 
-\trace\ Y( I_m - \sum_{ij} X_{ij} B_i^\adj B_j)
\\
&=&
\sum_{ij}\ X_{ij}(-W_{ji}-Z_{ji} + \trace\ YC_{ji}) - \trace\ Y
\eea
\eeq
where $Z=Z^\adj\in\Cbfnn$ and $Y=Y^\adj\in\Cbfmm$ are Lagrange
multipliers associated with the (Hermitian) inequality and equality
constraints, respectively. The {\em Lagrange dual function} is then,
\beq[eq:lag dual]
\bea{rcl}
g(Z,Y) 
&=& 
\inf_X\ L(X,Z,Y)
\\
&=&
\left\{
\bea{ll}
-\trace\ Y
&
Z_{ji} = \trace\ YC_{ij} - W_{ji}
\\
-\infty & \mbox{otherwise}
\eea
\right.
\eea
\eeq
For any $Y$ and $Z\geq 0$, $g(Z,Y)$ yields a lower bound on the optimal
objective $-\trace\ X^\opt W$. The largest lower bound from this dual
function is then $\max\set{g(Z,Y)}{Z\geq 0}$. Eliminating $Z$, this
can be written equivalently as,
\beq[eq:dual opt]
\bea{ll}
\mbox{minimize}
&
\trace\ Y
\\
\mbox{subject to}
&
K(Y) - W \geq 0,
\;\;
K_{ij}(Y) = \trace\ YC_{ij}
\eea
\eeq
with optimization variable $Y=Y^\dag\in\Cbfmm$. This is precisely the
result in \refeq{dual sdp}. Because the problem is strictly convex, the
dual optimal objective is equal to the primal optimal objective as
stated in the first line of \refeq{optcon}.  The {\em complementary
slackness} condition gives the second line in \refeq{optcon}.

\end{appendix}
\end{document}